\documentclass[iop,apjl]{emulateapj}
\usepackage{times}
\usepackage{epsfig}
\usepackage{amsmath, amsthm, amssymb}
\usepackage[plainpages=false, colorlinks=true, anchorcolor=blue, linkcolor=blue, citecolor=blue, bookmarks=false, urlcolor=blue]
{hyperref}
\usepackage{color}
\usepackage{microtype}
\usepackage{float}
\usepackage{verbatim}
\usepackage{wrapfig}

\newcommand{\beq}{\begin{equation}}
\newcommand{\eeq}{\end{equation}}
\newcommand{\bea}{\begin{eqnarray}}
\newcommand{\eea}{\end{eqnarray}}
\newcommand{\Msun}{$M_\sun$}

\newcommand{\MESA}{{\tt MESA}}
\newcommand{\FLASH}{{\tt FLASH}}

\newcommand{\subdate}{2015 March 6}
\newcommand{\shortauth}{Couch et al.}
\newcommand{\slugcom}{Submitted to ApJL on \subdate}
\slugcomment{\slugcom}

\lefthead{\sc \footnotesize \slugcom \hfill \shortauth}
\righthead{\sc \footnotesize \slugcom \hfill \shortauth}

\begin{document}

\title{The Three Dimensional Evolution to Core Collapse of a Massive Star}

\author{Sean M. Couch\altaffilmark{1}}
\author{Emmanouil Chatzopoulos\altaffilmark{2,6}}
\author{W. David Arnett\altaffilmark{3,4}}
\author{F.X. Timmes\altaffilmark{5}}

\affil{ 
  \altaffilmark{1}{TAPIR, Walter Burke Institute for Theoretical Physics, California Institute of Technology, Pasadena, CA 91125, USA;
    \href{mailto:smc@tapir.caltech.edu}{smc@tapir.caltech.edu}}\\ 
  \altaffilmark{2}{Flash Center for Computational Science, Department of Astronomy \& Astrophysics, University of Chicago, Chicago, IL 60637, USA}\\
  \altaffilmark{3}{Steward Observatory, University of Arizona, Tucson, AZ 85721, USA}\\
  \altaffilmark{4}{Aspen Center for Physics, Aspen, CO 81611, USA}\\
  \altaffilmark{5}{School of Earth and Space Exploration, Arizona State University, Tempe, AZ 85287, USA}
}
\altaffiltext{6}{Enrico Fermi Fellow}

\begin{abstract}

We present the first three dimensional (3D) simulation of the final minutes of iron core growth in a massive star, up to and including the point of core gravitational instability and collapse.
We capture the development of strong convection driven by violent Si burning in the shell surrounding the iron core.
This convective burning builds the iron core to its critical mass and collapse ensues, driven by electron capture and photodisintegration.
The non-spherical structure and motion generated by 3D convection is substantial at the point of collapse, with convective speeds of several hundreds of km s$^{-1}$.
We examine the impact of such physically-realistic 3D initial conditions on the core-collapse supernova mechanism using 3D simulations including multispecies neutrino leakage and find that the enhanced post-shock turbulence resulting from 3D progenitor structure aids successful explosions.
We conclude that non-spherical progenitor structure should not be ignored, and should have a significant and favorable impact on the likelihood for neutrino-driven explosions.
In order to make simulating the 3D collapse of an iron core feasible, we were forced to make approximations to the nuclear network making this effort only a first step toward accurate, self-consistent 3D stellar evolution models of the end states of massive stars.


\end{abstract}

\section{Introduction}
\label{sec:intro}
\setcounter{footnote}{6}

Real stars are not truly spherically-symmetric.
This is especially true for the interiors of massive stars at the end of their lives.
As massive stars approach core collapse, the equation of state becomes softer, cooling by neutrino emission drives nuclear burning ever more vigorously, and convective velocities increase. 
Nuclear burning couples to turbulent convection so that fuel is consumed in chaotic bursts. 
Core burning and thick shell burning are dominated by large scale modes of flow, which are of such low order that they do not cancel to a smooth spherical behavior.
In 1D stellar evolution codes, convective mixing and energy transport is modeled using mixing-length theory (MLT), which is tuned to reproduce the solar photosphere \citep{Asplund:2009}.
3D simulations of turbulent convection, however, demonstrate that MLT gives flawed representations of stellar convection, especially during the late burning stages \citep{Arnett:2015}, and two-dimensional (2D) simulations of the late stages of massive stellar evolution have shown that the Si burning in the shell surrounding the iron core is strong and violent, generating large-scale fluctuations in the core of the star that will be present at the point of core collapse \citep{{Bazan:1994fi},{Arnett:2011ga}}.

The state-of-the art in core-collapse supernova (CCSN) progenitor models \citep[e.g.,][]{{Woosley:2002ck}, {Woosley:2007bd}} is still 1D, and so simulations of the CCSN mechanism have all but exclusively used 1D initial conditions (ICs).
Recent exploration of the impact of multidimensional progenitor structure on the CCSN mechanism \citep{Couch:2013bl, Couch:2015gr, {Fernandez:2014if}, Mueller:2014wr} has shown that physically-motivated, yet highly parameterized, non-spherical structure in otherwise 1D progenitor models can have a qualitative and generally favorable impact on the CCSN mechanism.
Clearly, there is a need for realistic 3D progenitor models that  self-consistently capture the turbulent nuclear burning.

\begin{figure*}
  \centering
  \begin{tabular}{ccc}
    \includegraphics[height=2.6in]{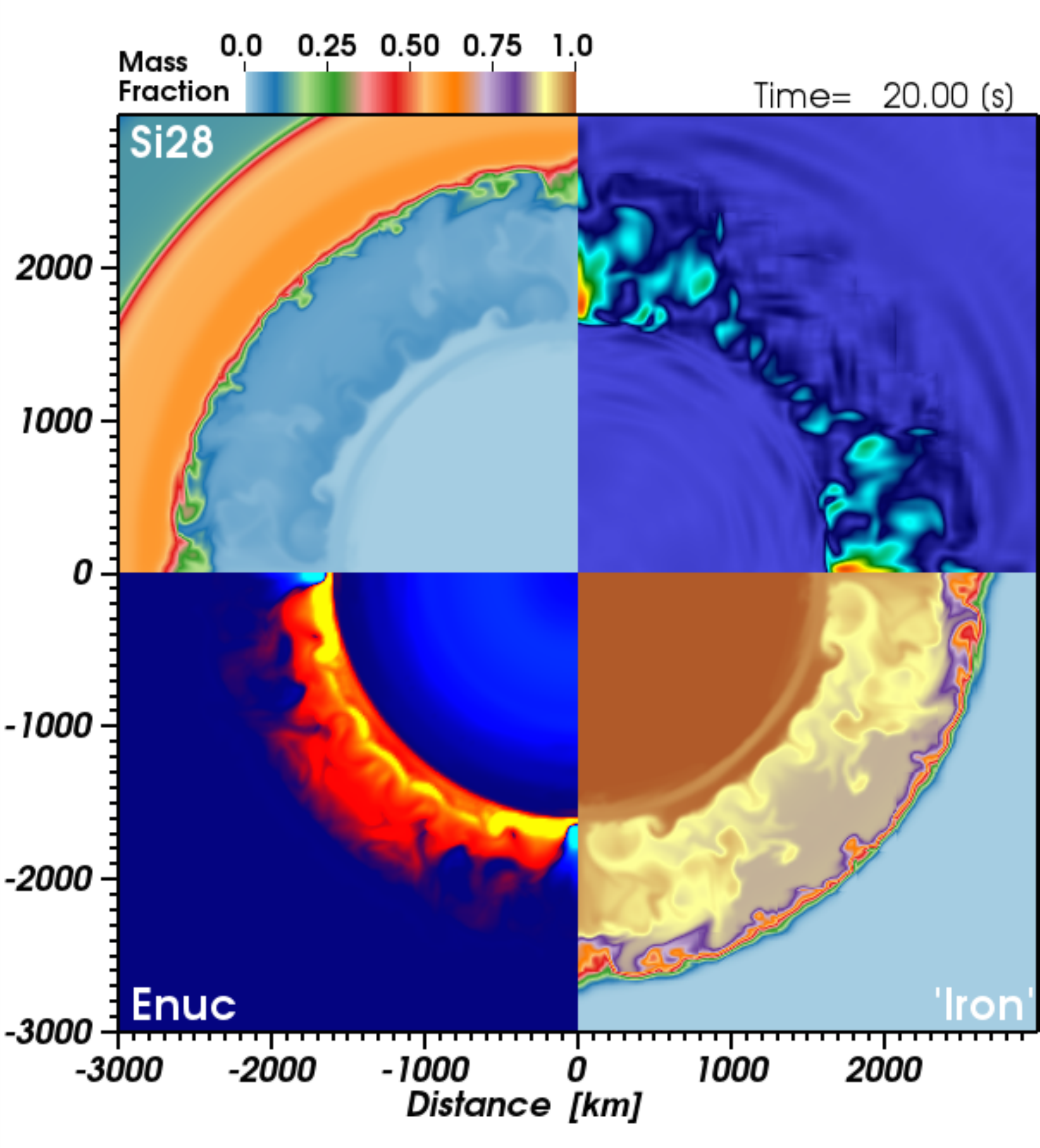} & \hspace{-0.17in}
    \includegraphics[height=2.6in]{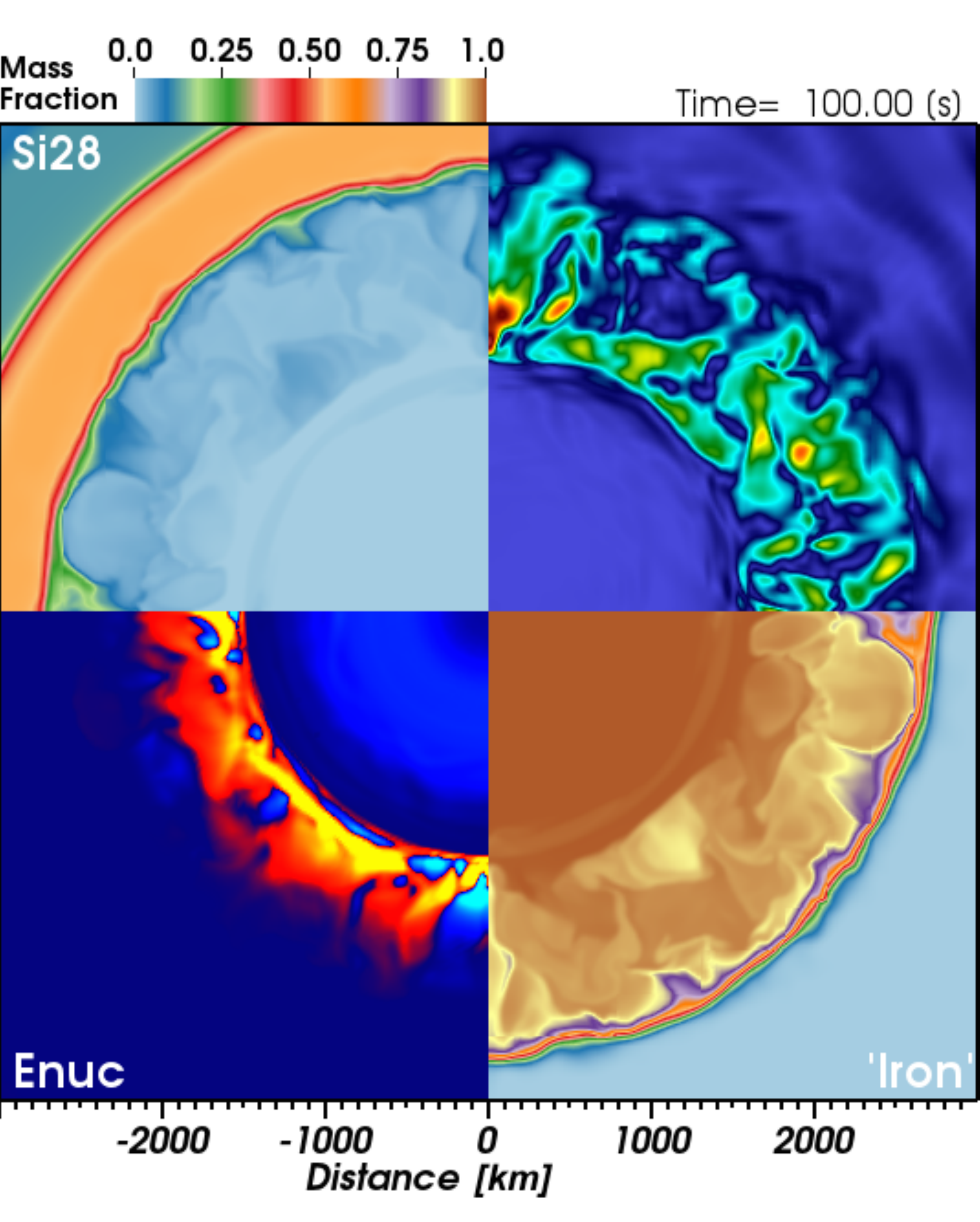} & \hspace{-0.17in}
    \includegraphics[height=2.6in]{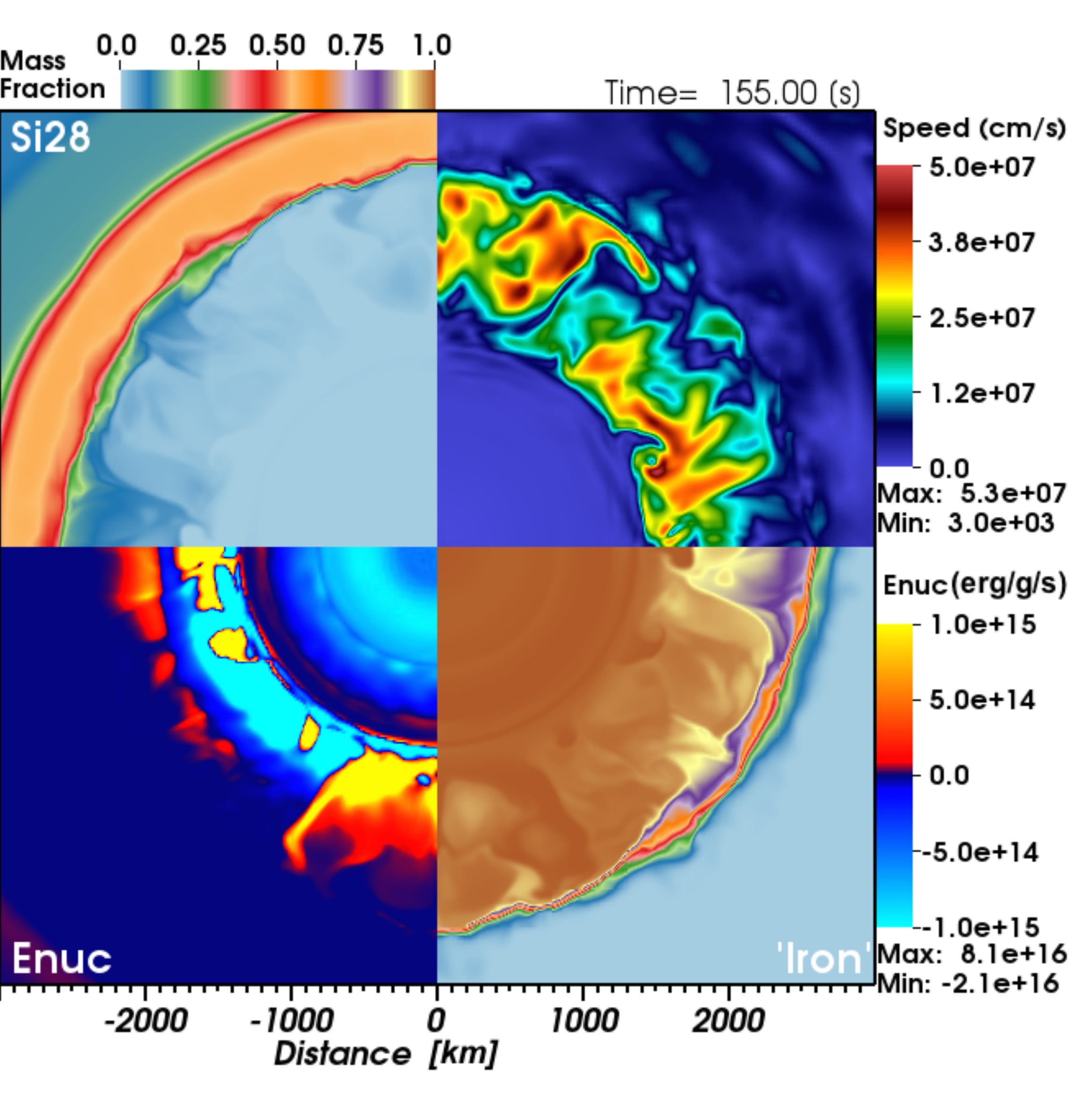}
  \end{tabular}
  \begin{tabular}{cc}
    \includegraphics[height=2.25in]{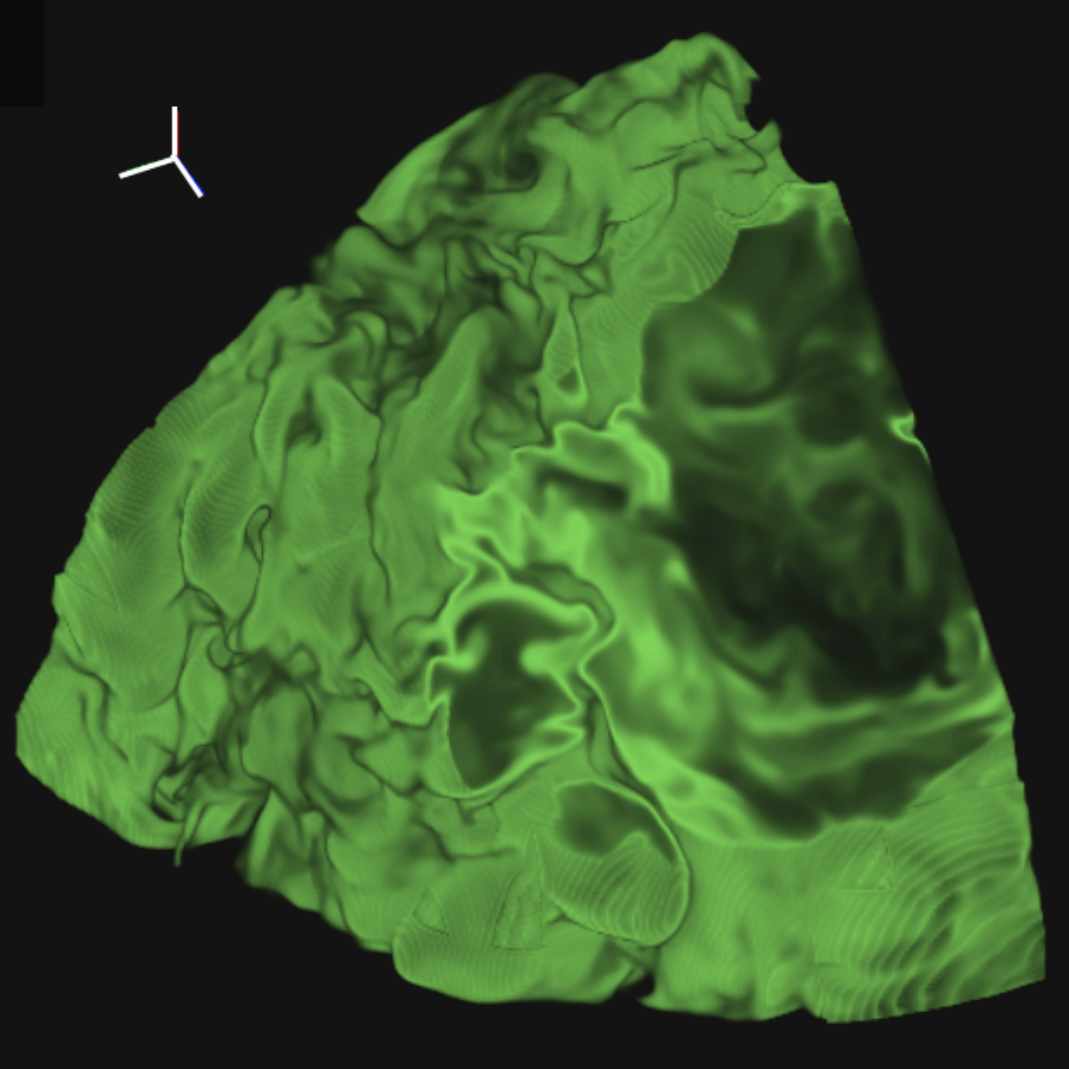} \hspace{-0.1in}
    \includegraphics[height=2.25in]{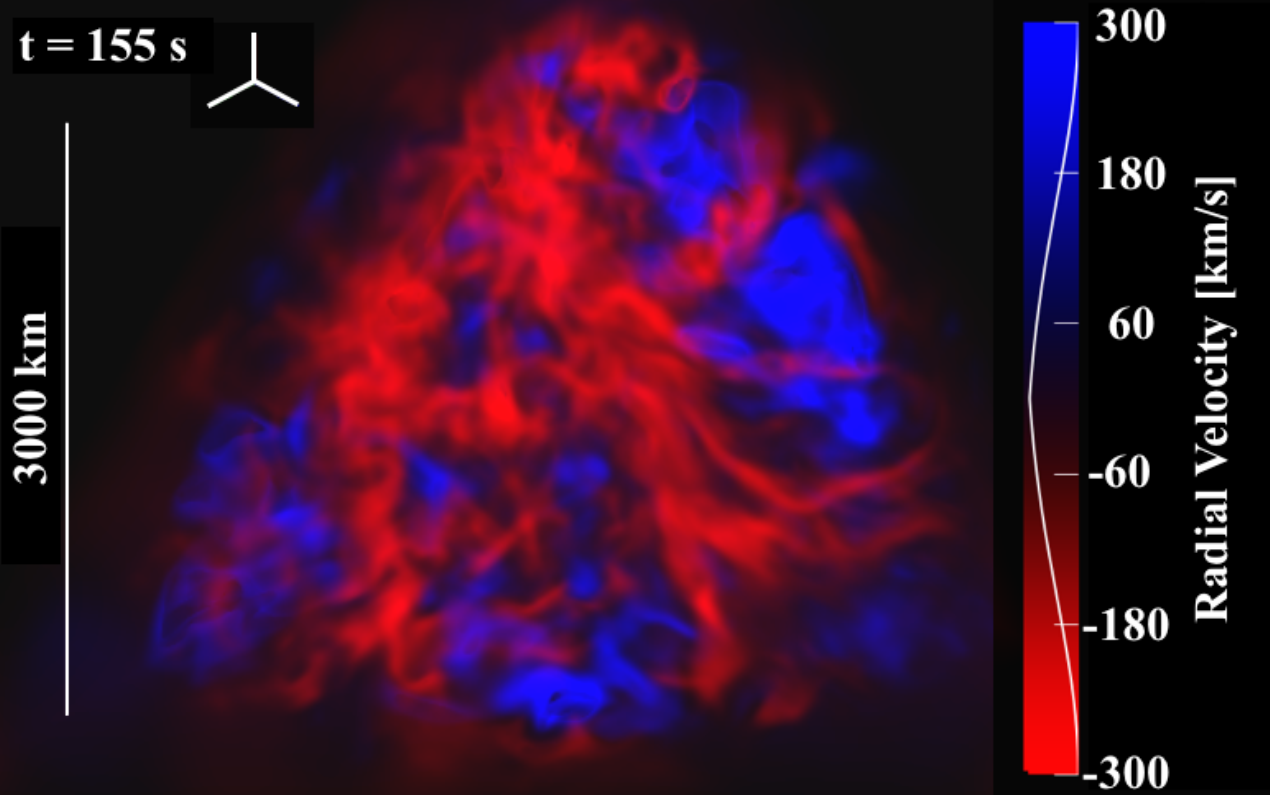}
  \end{tabular}
  \caption{
    Visualizations of the 3D progenitor evolution simulation.
    The top row displays pseudocolor slices of the $^{28}$Si mass fraction (top left), flow speed (top right), total mass fraction of iron group nuclei (bottom right), and specific nuclear energy generation rate (bottom left).
    The separate panels show different times since the start of the 3D simulation: 20 s (left), 100 s (middle), and 155 s (right).
    This final time is about 5 s before gravitational core collapse (see Figure \ref{fig:history}).
    The bottom row shows volume renderings of the surface where the `iron' mass fraction is 0.95 (left) and of the radial velocity (right) both at 155 s of 3D evolution.
  }
  \label{fig:viz}
\end{figure*}

In this Letter, we present the first 3D simulation of the late stages of iron core growth via strong Si shell burning through the moment of core collapse.
We concentrate on the final three minutes of Si burning in a 15 \Msun~star evolved initially in 1D.
This time scale captures several ($\sim$8) eddy turnover times of the convection in the Si shell and the synthesis of 0.2 \Msun~of iron.
Gravitational core instability is attained when the iron core reaches its effective Chandrasekhar mass, which depends in general on the core's electron fraction and entropy \citep{Baron:1990ik}.
We find the convection driven by Si shell burning to be strong with vigorous fluctuations. 
The turbulent speeds reach several hundred $\mathrm{km}\,\mathrm{s}^{-1}$, with the largest eddies being roughly the full width of the Si shell.
In order to assess the impact of realistic, physically self-consistent 3D progenitor structure on the CCSN mechanism, we follow the collapse of the 3D progenitor through core bounce and shock revival using approximate neutrino transport methods.
Compared to spherically-symmetric ICs, the presence of realistic 3D structure results in more favorable conditions for CCSN shock revival and robust explosion.
We find that this is due principally to stronger post-shock turbulence, which provides a greater effective turbulent pressure that aids shock expansion \citep{Murphy:2013eg, Couch:2015gr}.
Turbulence in the context of the CCSN mechanism is generated by neutrino-driven buoyant convection \citep{Murphy:2011ci, Abdikamalov:2014tc} and aspherical shock motion \citep{Endeve:2012ht}.
Progenitor asphericity introduces finite amplitude perturbations that enhance the growth rate of the instabilities that drive turbulence \citep[e.g.][]{Foglizzo:2006js}.
This first result shows that the final minutes of massive stellar evolution can, and should, be simulated in 3D.

The study of the CCSN mechanism with high-fidelity 3D simulations is still in its infancy.
Early results, however, indicate that progenitors that explode successfully in 2D \citep[e.g.,][]{Muller:2012gd, {Bruenn:2014wh}} may not in 3D \citep[e.g.,][]{Tamborra:2014do}.
There are multiple physical and numerical reasons why we should expect 2D simulations to be {\it artificially} prone to explosion \citep{{Hanke:2012dx}, {Couch:2013fh}, {Couch:2014fl}, {Takiwaki:2014hm}}, though the most important is likely the {\it inverse} turbulent energy cascade found in 2D simulations \citep[][and references therein]{Couch:2015gr}.
Observations indicate that massive stars explode successfully as CCSNe \citep[e.g.,][]{Smartt:2009kr}. 
The difficulty in obtaining explosions for massive stars in 3D simulations may indicate that something crucial is missing in our theory of the CCSN mechanism.
More realistic, 3D progenitor models will undoubtedly have an important impact on the CCSN mechanism and may be crucial to obtaining robust explosions. 

We proceed with a discussion of our simulation approach in Section \ref{sec:methods}.
We describe the 3D evolution during the final three minutes prior to core collapse of a massive star in Section \ref{sec:progen}.
In Section \ref{sec:ccsn}, we examine the impact of 3D progenitor structure on the CCSN mechanism.
Finally, we discuss the implications of our results and conclude in Section \ref{sec:conclusions}.

\section{Methods}
\label{sec:methods}

Our simulation of stellar core collapse proceeds in two steps.
First, we evolve a non-magnetic, non-rotating $15\,M_\odot$ star in 1D with the Modules for Experiments in Stellar Astrophysics \citep[\MESA; version 6794;][]{Paxton:2010jf, Paxton:2013km} to the point of iron core collapse, when the infall velocity of the outer core reaches $\sim$1000 km s$^{-1}$.
At a point prior to collapse, during quasi-hydrostatic Si shell burning when the iron core mass is around 1.3 $M_\odot$, we map the 1D \MESA~model into 3D using the \FLASH~simulation framework \citep{{Fryxell:2000em}, Dubey:2009wz}.

For the \MESA~models,\footnote{Complete \MESA~parameters inlist available at \url{http://flash.uchicago.edu/~smc/progen3d}.} we use the ``Helmholtz'' equation of state \citep[EOS,][]{Timmes:2000wm}.
The nuclear reaction network was automatically extended during the evolution
starting from a basic 8-isotope network and reaching a 21-isotope network
(``approx21'') by the end of the calculation. 
Standard mass-loss prescriptions were adopted \citep{Vink:2001,Glebbeek:2009}.
We use the Ledoux criterion for convection including semi-convection, thermohaline mixing, and overshoot. 
We use a resolution parameter for the \MESA~calculations of 0.6, corresponding to 3646 zones in 1D at the time of mapping to 3D. 

For the 3D \FLASH~simulation, we also employ the Helmholtz EOS and the same 21-isotope network, newly implemented in \FLASH~from the standalone public network.\footnote{\url{http://cococubed.asu.edu/code_pages/burn\_helium.shtml}}
The approx21 network includes a very approximate treatment of heavy element neutronization important for the near-collapse isotopic evolution of the iron core.
Ideally, one would use a network with sufficient number of isotopes ($\sim$60 -- 100) 
 to treat the core neutronization and URCA cooling directly.
For simplicity, speed, and in order to maintain direct network equality between the 1D \MESA~models and the 3D \FLASH~simulations we use the reduced network.

We include the complete iron core in the 3D domain.
The tools of stellar evolution model convective burning in essentially an average sense, neglecting multidimensionality and highly-dynamic fluctuations.
This gives rise to a practical problem for multidimensional simulation of stars. 
Convective progenitor models do not have consistent 3D turbulent flow, and when they are mapped onto a 3D grid as an initial state for hydrodynamic simulations, they always pass through a transient state during which a turbulent flow develops. 
This in not a problem of mathematics (the mapping) but one of physics (the turbulence). 
The 1D stellar models cannot provide the 3D information for the turbulent fluctuations (both amplitudes and phases). 
There are no known successful attempts to fake it; transients happen. 
Ideally the mapping from 1D to 3D should occur earlier in the evolution such that any initial transients are negligibly small.

In order to quell such initial transients due to the mapping from 1D to 3D, which typically manifest as strong radial waves as the star settles onto the new domain, we employ the hydrostatic initialization approach of \citet{Zingale:2002kk} rather than initial damping \citep[e.g.,][]{Arnett:1994go}.
In this approach, the density profile is adjusted slightly while keeping the pressure profile fixed such that the equation of hydrostatic equilibrium is satisfied exactly throughout the model.
This procedure is closed using the EOS.
Nevertheless, some initial transient motion persists during the first $\sim$20 s of the 3D evolution.
This erroneous expansion can effectively ``de-evolve'' the star, pushing the stellar structure into a state resembling earlier phases of the evolution.
This results in the requirement for a prohibitively long period of simulation time to reach collapse.
In order to overcome this lack of realism in the initial model, we enhance the parameterized rate of the neutronization reaction that converts $^{56}$Fe to $^{56}$Cr, thus enhancing the cooling of the inner part of the iron core to compensate for the inefficient rate of outward energy transport.
The rate of this reaction was enhanced above the fiducial value by a factor of 50.
This approach, though admittedly non-ideal, effectively damps the initial core expansion allowing us to reach iron core collapse in 3D in a reasonable amount of simulation time.

For the 3D \FLASH~simulation, we use Cartesian coordinates coupled with adaptive mesh refinement (AMR) in one octant of the full 3D sphere.
The finest grid spacing, using 8 levels of refinement, is 16 km.
The refinement level is reduced as a function of radius, with the first reduction occurring around a spherical radius of 2500 km, beyond the Si-burning shell.
The entire domain is 100,000 km on a side.
Along the octant symmetry planes we use reflecting boundary conditions while at the outer extents of the domain we use zero-gradient boundary conditions.
For solving the hydrodynamics, we use directionally-unsplit PPM (without contact steepening) and the HLLC Riemann solver.
The unsplit PPM solver in \FLASH~does not utilize the ``consistant multifluid advection'' scheme of \citet{Plewa:1999tn}.
We include self-gravity assuming a spherically-symmetric (monopole) gravitational potential.
Due to the operator splitting between the hydrodynamics and the nuclear network, we find that, in order to maintain adequate coupling between the burning and the hydrodynamics, it is critical to limit the size of the time step so that the internal energy in any one zone changes by no more than 1\% during the course of a single step.

\section{Three Dimensional Collapse of an Iron Core}
\label{sec:progen}

\begin{figure}
  \centering
  \includegraphics[width=3.4in]{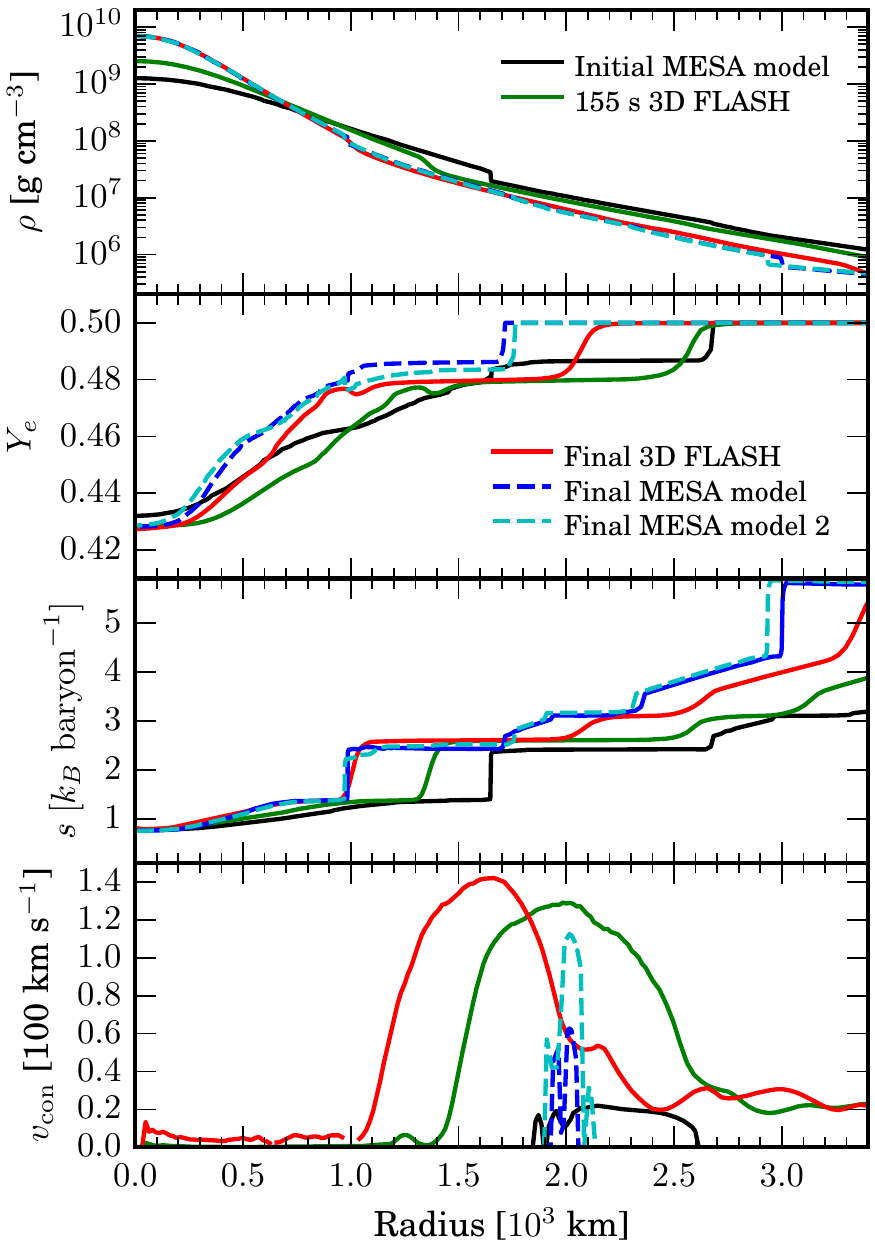}
  \caption{
    Spherically-averaged density (top), electron fraction (middle), and convective speed, $v_{\rm con} = \langle \lvert v_r - \langle v_r \rangle \rvert \rangle$, (bottom) for the 3D progenitor simulation at three different 3D evolution times: 0 s (initial MESA model, black lines), 155 s (green lines), and 160 s (collapsing model, red lines).
  }
  \label{fig:models}
\end{figure}

We follow the violent Si shell burning and build up of the iron core mass in 3D for $\sim$160 s.
This single simulation required approximately 350,000 core-hours on {\it Stampede} at TACC on 1024 cores.
Visualizations of the 3D progenitor burning simulation are shown in Figure \ref{fig:viz}, where we show slice plots of key quantities along with volume renderings of the iron core and the radial component of the velocity in the Si-burning shell.\footnote{Movies of these visualizations may be viewed at \url{http://flash.uchicago.edu/~smc/progen3d}.}
In Figure \ref{fig:models}, we show spherically-averaged radial profiles of the density, $\rho$, electron fraction, $Y_e$, specific entropy, $s$, and convective velocity from the 3D \FLASH~simulation at three times: the transition from the 1D \MESA~model to 3D; 5 s prior to collapse; and the point of collapse.
Angle averages, $\langle ... \rangle$, are taken over spherical shells and $v_r$ is the radial velocity component.
Evident from Figure \ref{fig:models} is that the convective speeds near collapse are typically $>$100 km s$^{-1}$.
This is slightly larger than the comparable speeds found in the O-burning shell \citep{vmam13}. Because the nuclear burning is balanced on average by turbulent dissipation, the energy generation rate is related to the average velocity and the depth of the convection zone by $\epsilon \sim v^3/\ell$ \citep{Arnett:2009}. 
This is, however, the {\it average} convective speed, and fluctuations increase the peak speeds.  We see from Figure \ref{fig:viz} that the peak speeds in the Si-burning shell can be {\it several} hundred km s$^{-1}$, reaching speeds near collapse of $\sim$500 km s$^{-1}$. This is not negligible when compared to nominal infall speeds for core collapse initial models ($\sim$1000 km s$^{-1}$).
The speed of the convection increases as collapse approaches and the core contracts.

In Figure \ref{fig:models} we also show final 1D \MESA~models at the point of collapse considering two different scenarios: one in which the neutronization reaction rate is enhanced by the same amount as in the 3D \FLASH~simulation (blue dashed lines) and the other in which we do not enhance this reaction rate above the fiducial value found in \MESA's approx21 network (cyan dashed lines).
Stellar collapse is highly dynamic and the model profiles change rapidly once gravitational instability sets in. 
Thus, for the sake of fair comparison, we consider all models at the point when the central densities have reached the same value as that of the final 3D \FLASH~simulation ($7\times10^9$ g cm$^{-3}$).
Varying the rate of neutronization has very little effect on the 1D \MESA~profiles at the point of collapse.
It does, however, dramatically change the time it takes to reach collapse.
For the fiducial neutronization rate, the \MESA~model takes 1000 s to reach collapse from the model we used for the 3D ICs (i.e., the black lines in Figure \ref{fig:models}).
With the neutronization rate enhanced as we have done for the 3D \FLASH~simulation, the \MESA~model requires just 40 s to reach collapse.
This is faster even than the \FLASH~simulation because the \MESA~model, obviously, does not experience the initial transient pulsation that the 3D simulation must go through until vigorous convection is established.
The final iron core masses of all models are similar, though correlated with the time it takes to reach collapse: 1.46 \Msun~for the enhanced-rate \MESA~model, 1.51 \Msun~for the fiducial-rate \MESA~model, and 1.50 \Msun~for the 3D \FLASH~simulation.

\begin{figure}
  \centering
  \includegraphics[width=3.4in]{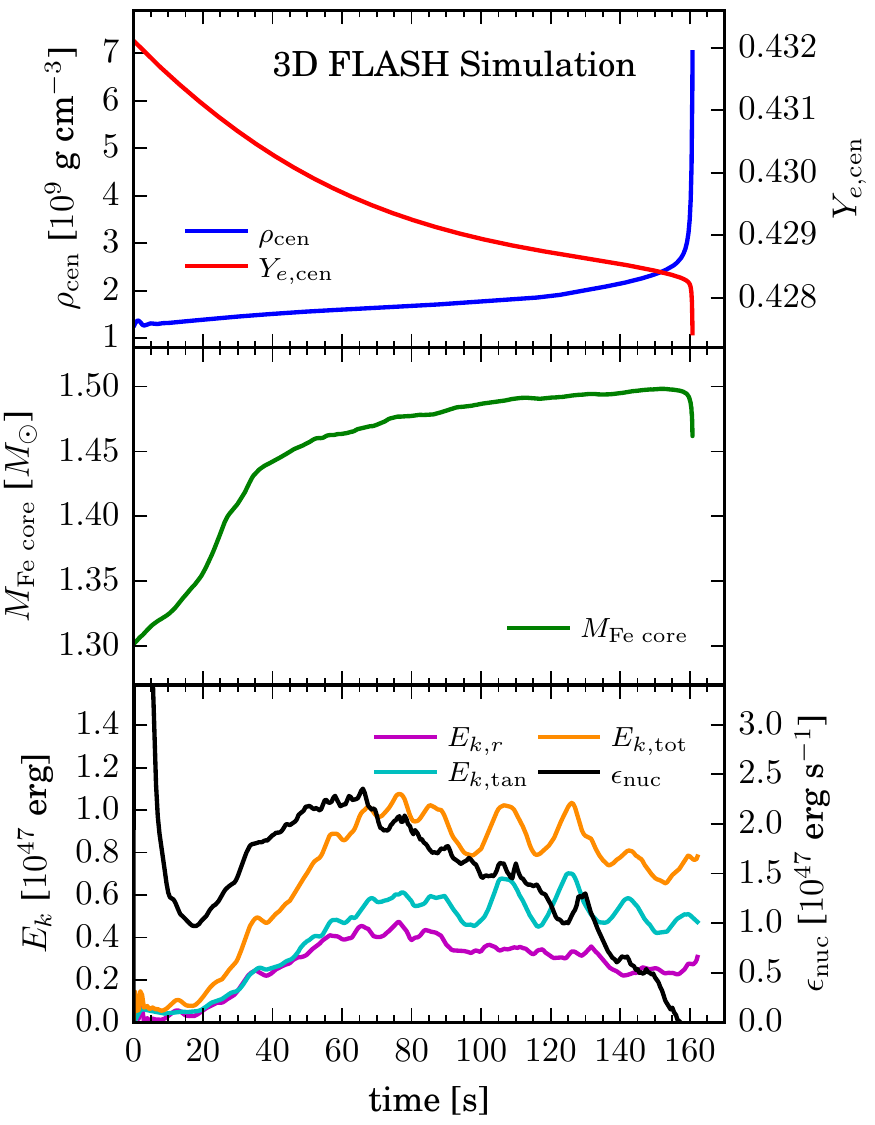}
  \caption{
    Time histories from the 3D progenitor simulation of the central density and central electron fraction (top panel), mass of the iron core (middle panel), and the total kinetic energies in the radial and tangential directions along with the net nuclear energy release rate in the Si-burning shell (bottom panel).    
  }
  \label{fig:history}
\end{figure}  

Figure \ref{fig:history} shows the time evolution of the central density, central electron fraction, the mass of the iron core, the total kinetic energy, and the net nuclear energy release rate in the Si-burning shell for the 3D \FLASH~simulation.
We define the Si-burning shell as the region of the star where both the iron and silicon mass fractions exceed 0.001.
During the simulated time, the iron core mass grows from 1.3 \Msun to 1.5 \Msun, while the central $Y_e$ decreases from 0.432 to $\sim$0.428 at collapse.
The initial transient wave during the first 20 s is evident as an excess of radial kinetic energy in the bottom panel of Figure \ref{fig:history}.
After the brief transient, the kinetic energy is dominated by convective motion in the Si-burning shell until the collapse begins, around 160 s.
The growth in the kinetic energy corresponds directly with a positive increase in the net nuclear energy release rate, which is negative because it is dominated by strong cooling from the central core.
We expect that the convective kinetic energy is not unduly influenced by initial transient behavior because it reaches a near steady-state long before the end of the 3D simulation, and because we have simulated several convective eddy turnover times.
Assuming an average convective speed of 100 km s$^{-1}$, and a width of the Si-burning convective region of $\sim$1000 km (see Figure \ref{fig:models}), the turnover time is approximately $\Delta t_{\rm eddy} \sim 2 \Delta r / v_{\rm con} \sim 20$ s.
Driven convection/turbulence reaches a quasi-steady-state on roughly only a few turnover times \citep[e.g.,][]{{Meakin:2007dj}, Radice:2015ty}.

\begin{figure}
  \centering
  \includegraphics[width=3.4in]{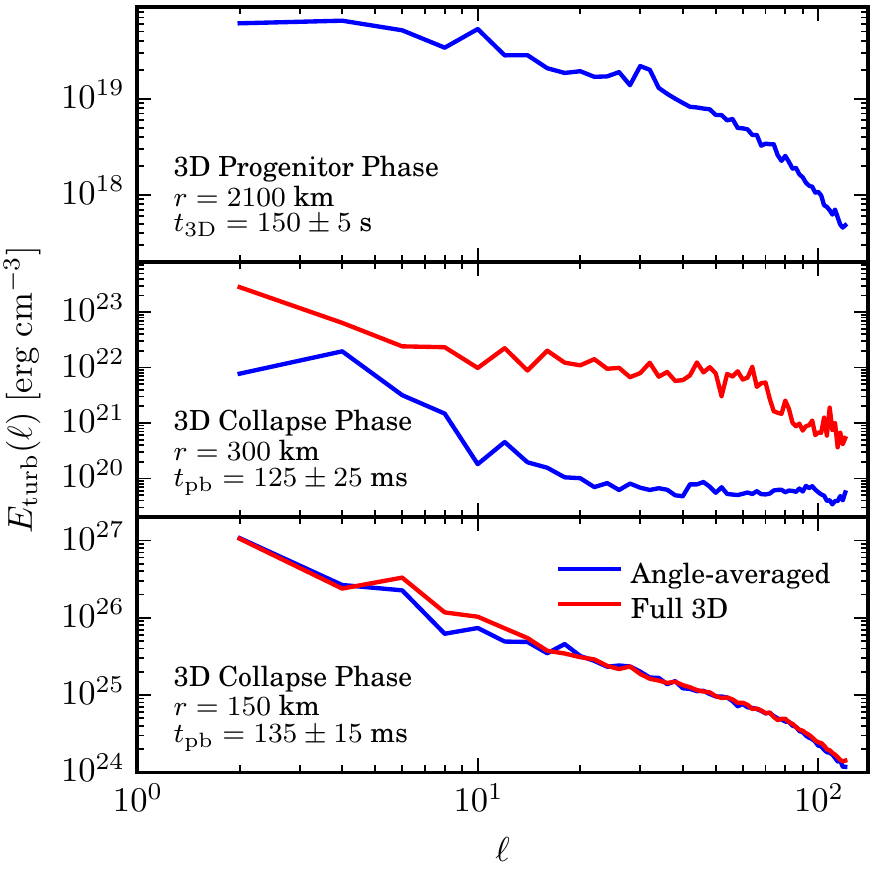}
  \caption{
    Turbulent kinetic energy spectra in spherical harmonic basis. 
    The top panel shows the turbulent energy spectrum for the 3D progenitor simulation averaged over the 10 s prior to the start of core collapse averaged over a spherical shell centered on 2100 km (i.e., the Si-burning convective region).
    The middle panel show the turbulent energy spectrum during the collapse phase around a post-bounce time $t_{\rm tb} = 125$ ms, averaged over a shell centered on a radius of 300 km, i.e., ahead of the shock in the accretion flow.
    Two different cases are shown: the 3D ICs (red) and the 1D ICs based on spherically-averaging the 3D progenitor simulation (blue).
    This spectrum quantifies the strength of pre-shock turbulent fluctuations that then influence the post-shock turbulence once accreted through the shock.
    For the angle-averaged case (blue lines), this gives an estimate of the perturbations introduced by the Cartesian AMR grid.
    The bottom panel shows the turbulent energy spectra also during the collapse phase, around 135 ms post-bounce, but for a shell situated in the gain region.
  }
  \label{fig:spectra}
\end{figure}

Figure \ref{fig:viz} also shows the initiation of convection from the initially spherically-symmetric model (right-most panel of top row). 
As in the 2D simulations of \citet{{Arnett:2011ga}}, the Si shell burning is inhomogeneous with alternating regions of net cooling by neutrinos and net heating by nuclear burning.
The iron core shows overall contraction while the Si shell is depleted by the burning.
The bottom row of Figure \ref{fig:viz} shows that, near collapse, the surface of the iron core is significantly distorted from spherical symmetry. 
At this time, the average-weighted standard deviation of the radial coordinate of the surface of the iron core is 9.5\%.
The radial velocity also shows that the velocity fluctuations are large in both scale and amplitude.
We quantify the spatial scale (as well as strength) of the turbulent convection in the Si shell by measuring the turbulent kinetic energy power spectrum in spherical harmonic basis \citep[for details on how this is computed, see][]{Couch:2014fl}.
For the 3D progenitor simulation near collapse, this is shown in the top panel of Figure \ref{fig:spectra}.
The turbulent energy spectra peak at quite small $\ell$ (large scale), around $\ell \approx 4$.
This implies that the largest eddies, which carry the bulk of the convective/turbulent kinetic energy, have roughly the same radial extent as the convective shell, precisely as observed in Figure \ref{fig:viz}.
We have also analyzed the convection using the vector spherical harmonic framework of \citet{Chatzopoulos:2014gj} and find that the vector spectra also peak around $\ell \approx 5$.

The final stages of nuclear burning in massive stars can excite waves that could have important implications for mass loss and angular momentum transport just prior to core collapse \citep{{Meakin:2007dj},{Quataert:2012jg}, {Shiode:2014il}, Fuller:2015}.
We find that the Si shell burning in the minutes before collapse drives non-spherical gravity waves that propagate both outward away from the core and {\it inward} into the iron core.
Previous multidimensional studies of late-stage stellar burning \citep[e.g.][]{Bazan:1994fi, {Meakin:2007dj}, {Arnett:2011ga}} saw such phenomena but lacked fully dynamic inner cores and were not able to assess them accurately.
Such waves persist throughout the Si burning phase and are present at the point of collapse, as evident in the spherically-averaged convective velocity shown in Figure \ref{fig:models}.
The wave velocity amplitudes at the start of collapse are tens of km s$^{-1}$.
The contraction, and ultimately unstable collapse, of the iron core amplifies these waves \citep{Lai:2000hv} and they become quite substantial around the moment of core bounce.

\section{Impact on the Supernova Mechanism}
\label{sec:ccsn}

\begin{figure}
  \centering
  \includegraphics[width=3.4in]{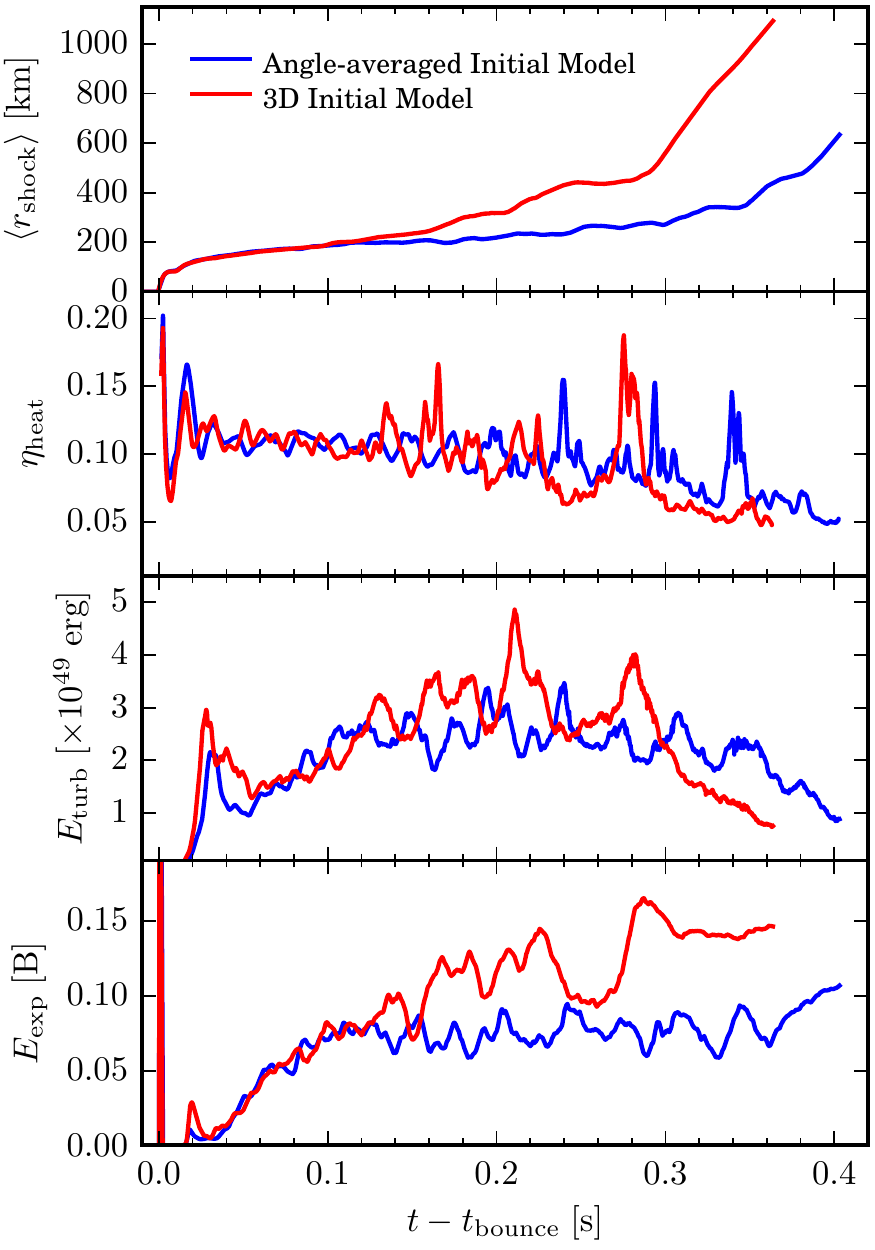}
  \caption{
    Results of core-collapse simulations comparing the full 3D ICs resulting from the 3D progenitor simulation (red lines) to the 3D simulation using angle-averaged ICs constructed from the 3D progenitor simulation.
    The top panel shows the average shock radius, the second panel shows the neutrino heating efficiency, the third panel shows the total turbulent kinetic energy in the gain region, and the bottom panel shows the diagnostic explosion energy (see text for relevant definitions).
  }
  \label{fig:ccsn}
\end{figure}

Our primary interest in simulating the final stages of stellar evolution in 3D is to assess whether realistic multidimensional progenitor structure has a significant impact on the CCSN mechanism.
In order to achieve the 3D simulation of the final minutes of a massive star's life up to the point of core collapse we had to make certain approximations: (1) use of a simplified nuclear reaction network and adjusted electron capture rates, (2) simulating only the final moments of evolution after O shell burning has ended (for this model), and (3) use of a 1D MLT initial model.
In this section we present a brief examination of the impact of such idealized but fully 3D ICs on the CCSN mechanism.

We carry out two 3D Newtonian simulations of collapse, bounce, and shock-revival including approximate neutrino physics.
For one simulation, we use the full 3D progenitor simulation continued from the moment of collapse ($t_{\rm 3D} \approx 160$ s).
In the other, we angle-average the final 3D progenitor model, washing away all non-radial velocities and non-spherical structure, to produce 1D ICs which are used in an otherwise fully 3D simulation. 
Our simulation approach is essentially identical to that of \citet{Couch:2014fl} and \citet{Couch:2015gr}, with the only major difference being that our 3D simulations here are carried out only in one octant, rather than the full star.
We use Cartesian coordinates with adaptive mesh refinement, yielding a finest grid spacing of $\sim$0.5 km and an effective angular resolution of $\sim$0.5$^\circ$.
Neutrino effects are incorporated using a multispecies leakage scheme that includes charged current heating and pre-bounce deleptonization is approximated using the density-dependent approach of \citet{Liebendorfer:2005ft}.
We enhance the {\it local} charged current heating rates by 6\% in both cases in order to yield explosions.

Figure \ref{fig:ccsn} summarizes our investigation of the impact of 3D ICs on the CCSN mechanism.
The top panel of Figure \ref{fig:ccsn} shows the average shock radius of the full 3D ICs simulation (red lines) and the angle-averaged 1D ICs simulation (blue lines).
The average shock radii between the two simulations is nearly identical until around 100 ms after bounce.
This corresponds with the time at which the Si shell interface is accreted through the shock surface.
At this time, the shock in the full 3D ICs model begins to expand more rapidly than for the 1D ICs model, behavior that continues as both models transition to explosion.
The second panel of Figure \ref{fig:ccsn} shows the neutrino heating efficiency, $\eta_{\mathrm{heat}} = Q_{\rm net} (L_{\nu_e,\mathrm{gain}} + L_{\bar{\nu}_e,\mathrm{gain}})^{-1}$, where $Q_{\rm net}$ is the net charged current heating rate in the gain region, which is divided by the sum of the electron-type neutrino and antineutrino luminosities at the base of the gain layer.
We find essentially no difference in the neutrino heating efficiencies between the two simulations.
The same can be said also for the neutrino luminosities and the total gain-region heating rates.
Clearly, the cause for the divergence in the results between the 3D ICs and the 1D ICs is not due principally to differences in the neutrino heating or average gain region matter dwell times \citep[e.g.,][]{{Murphy:2011ci}, Murphy:2008ij, {Dolence:2013iw}}.
Our present results are consistent with the picture in which the differences are due primarily to different strengths of post-shock turbulence, and the attendant effective turbulent pressures that aid shock expansion \citep{Couch:2015gr}.
The third panel of Figure \ref{fig:ccsn} shows the total turbulent kinetic energies in the gain region for both simulations.
The 3D ICs simulation has greater turbulent energy after about 100 ms, following accretion of the Si interface.
There is also a larger ``burst'' of turbulent energy immediately post-bounce caused by the presence of aspherical, convection-generated waves in the inner part of the core right at bounce.

We also find differences in the turbulent kinetic energy spectra between the 3D ICs simulation and the 1D ICs simulation.
The bottom two panels of Figure \ref{fig:spectra} show the turbulent energy spectra from the two collapse simulations at different radii: $\sim$300 km (ahead of the shock) and $\sim$150 km (in the gain region).
The spectrum for the 1D ICs simulation for the region ahead of the shock, which should be essentially a spherically-symmetric accretion flow, quantifies the magnitude of the perturbations excited by our use of a Cartesian grid.
The turbulent energy in this region for the full 3D ICs is an order of magnitude or more greater than this at all scales (except $\ell = 4$).
This translates into greater turbulent energy in the gain region, as seen in the bottom panel of Figure \ref{fig:spectra}.
The 3D ICs result in greater turbulent energy on large scales, between about $\ell = 6-10$, precisely where it is most effective at aiding shock expansion \citep{{Hanke:2012dx}, Couch:2014fl, {Couch:2015gr}}.
The excess of power in the quadrupole $\ell = 2$ mode is likely due to our use of an octant domain.

The greater strength of turbulence excited by realistic 3D ICs also results in a greater diagnostic explosion energy, as seen in the bottom panel of Figure \ref{fig:ccsn}.
Here, the diagnostic explosion energy, $E_{\rm exp}$, is the total energy of all gravitationally unbound material with net positive radial velocity \citep[see definitions in, e.g.,][]{Muller:2012gd, {Bruenn:2014wh}}.
We note, however, that at the end of our simulations the explosion energies are not near their asymptotic final values, so caution should be used when interpreting the differences in the diagnostic explosion energies.

\section{Discussion and Conclusions}
\label{sec:conclusions}

We have carried out the first 3D simulation of the final growth of the iron core in a massive star up to the point of gravitational instability and collapse.
The violent Si burning in the shell surrounding the iron core drives large-scale, strong deviations from spherical symmetry that is not captured by 1D models.
We show that this has a positive impact on the favorability for explosion via the delayed neutrino heating mechanism for CCSNe.
We were forced to make a number of approximations, e.g., the use of a much reduced nuclear network, modifying electron capture rates, using MLT 1D initial models, simulating only one octant of the full star, and focussing on only one initial progenitor model.
Thus, crucial aspects of the final iron core, such as the electron fraction and entropy, may be affected.
Additionally, the presence of a Si-burning shell at the moment of core collapse may be progenitor model-dependent \citep[see, e.g.,][]{{Woosley:2002ck}}.

Compared to 2D simulations of Si shell burning \citep{Arnett:2011ga}, the maximum convective speeds we find are somewhat smaller, in agreement with 2D vs.\ 3D results found for O burning shells \citep{Meakin:2007dj}.
The speeds we find are smaller, also, than those assumed in the parameterized velocity fluctuations employed by \citet{Couch:2013bl}.
As a result, the impact we find on the CCSN mechanism may be less dramatic than that found by \citet{Couch:2013bl}, though it is still significant.
We show that the biggest impact of 3D ICs is the enhancement of the strength of post-shock turbulence.
Greater turbulence behind the shock leads to a greater effective turbulent pressure and, thus, more favorable conditions for shock expansion and explosion \citep{Murphy:2013eg, {Couch:2015gr}}.

Realistic 3D simulations of the convective nuclear burning in massive stars, comparable to those used here, also suggest modifications to the MLT algorithms used in stellar evolution calculations \citep{Arnett:2015}.
Here, we have focussed on the impact that 3D ICs have on the CCSN mechanism, showing that the breaking of spherical symmetry has a significant and positive impact on the likelihood for explosion.
This is a critical finding since robust neutrino-driven explosions have been notoriously difficult to achieve across the broad range of (spherically-symmetric) progenitors studied to-date \citep[see, e.g.,][]{Janka:2012cb}.
The first 3D simulations including detailed neutrino physics indicate that explosion may be harder to attain than for 2D \citep{Tamborra:2014do}.

The CCSN ``problem,'' i.e., the persistent difficulty in achieving robust explosions, has been with us for nearly a half century since the neutrino mechanism was introduced \citep{Colgate:1966cl,wda68}.
Most theoretical investigations of the CCSN mechanism have, however, employed spherically-symmetric, non-rotating, non-magnetic stars.
It is well known that all stars rotate to some degree, are endowed with magnetic fields, and are spherically-symmetric only in the average sense.
We have shown here that the detailed 3D structure of massive stellar cores, that is a natural and unavoidable result of the final stages of nuclear burning, has a big and important impact on the CCSN mechanism.
Additional investigation of the roles that rotation and magnetic fields play in both the pre-collapse evolution of massive stars and in the CCSN mechanism itself is needed, but taken together with realistic 3D progenitor structures, it is conceivable that attaining energetic explosions that reproduce the observable and statistical properties of CCSNe \citep{Clausen:2015} may not be so challenging.
That is, there might not be a CCSN ``problem'' per se, we may have just been using progenitor models that do not occur in nature, and given a set of progenitor models more representative of real massive stars, the ``problem'' may go away.

This work is a first step toward addressing these issues.
It serves to a large extent as a proof-of-principle indication that the final minutes of massive star evolution can and should be simulated in 3D. 
Doing so has a significant impact on the final structure of the core and on the CCSN mechanism.
There is significant progress that must still be made in accurately simulating CCSN progenitors, including better models for mixing during the long, evolutionary phases of the star's life, and inclusion of realistic treatments for rotation and magnetism.

\acknowledgements 

We thank J. Fuller, C. Graziani, C. Meakin, C. Ott, and M. Zingale for many valuable conversations.
SMC is supported by the National Science Foundation under award no. AST-1212170.
EC would like to thank the Enrico Fermi Institute for its support.
FXT thanks NASA for partial support under TCAN award NNX14AB53G.
This research was partially supported by the National Science Foundation under award no. AST-1107445 at the University of Arizona.
The software used in this work was in part developed by the DOE NNSA-ASC OASCR Flash Center at the University of Chicago.  
This research used computational resources at ALCF at ANL, which is supported by the Office of Science of the US Department of Energy under Contract No. DE-AC02-06CH11357, and at TACC under NSF XSEDE allocation TG-PHY100033.


\end{document}